\documentclass{tPRS2e}

\usepackage{epstopdf}
\usepackage{subfigure}
\usepackage{amsmath}
\usepackage{amssymb}
\usepackage{txfonts} 
\usepackage{booktabs} 
\usepackage{multirow}
\usepackage{tabularx}
\usepackage{array}
\usepackage[ruled,linesnumbered]{algorithm2e}
\usepackage{overpic}
\usepackage{pgfplots}
\usepackage{filecontents}
\usepackage{lineno}
\usepackage{url}
\begin{document}


\articletype{RESEARCH ARTICLE}

\title{Single item stochastic lot sizing problem considering capital flow and business overdraft}

\author{
\name{Zhen Chen\textsuperscript{a},
 Roberto Rossi\textsuperscript{b}, Ren-qian Zhang\textsuperscript{a}$^{\ast}$\thanks{$^\ast$Corresponding author. Email: zhangrenqian@buaa.edu.cn}}
\affil{\textsuperscript{a}School of Economics and Management, Beihang University, Beijing, China\\
\textsuperscript{b}Business School, University of Edinburgh, Edinburgh, UK}
\received{v5.0 released June 2015}
}

\maketitle
\begin{abstract}
This paper introduces capital flow to the single item stochastic lot sizing problem. A retailer can leverage business overdraft to deal with unexpected capital shortage, but needs to pay interest if its available balance goes below zero. A stochastic dynamic programming model maximizing expected final capital increment is formulated to solve the problem to optimality. We then investigate the performance of four controlling policies: ($R, Q$), ($R, S$), ($s, S$) and ($s$, $\overline{Q}$, $S$); for these policies, we adopt simulation-genetic algorithm to obtain approximate values of the controlling parameters. Finally, a simulation-optimization heuristic is also employed to solve this problem. 
Computational comparisons among these approaches show that policy $(s, S)$ and policy $(s, \overline{Q}, S)$ provide performance close to that of optimal solutions obtained by stochastic dynamic programming, while simulation-optimization heuristic offers advantages in terms of computational efficiency. Our numerical tests also show that capital availability as well as business overdraft interest rate can substantially affect the retailer's optimal lot sizing decisions.
\end{abstract}


\section{Introduction}

Capital shortage is a key factor affecting the growth of many small and medium enterprises (SMEs). In contrast to large companies, due to sales volume and operation scales, it is not easy for SMEs to obtain external capital in the form of loans or venture capital investment. In addition, SMEs usually lack the capital to absorb large losses. A survey of 531 businesses that went bankrupt during the calendar year 1998 in \cite{bradley2000lack} pointed out that inadequate financial planning was one of main reasons for their business failing. A report by \cite{coughtrie2009restructuring} showed shortage of capital accounted for 17\% of company bankruptcies in Australia in 2008.  \cite{elston2011financing22} found that 84\% of high-tech entrepreneurs in the US had experienced a shortage of capital at some time.  

Numerical experiments in \cite{zeballos2013single} demonstrate that the lack of access to short-term debt drastically inflates working capital requirements and lowers cash flows. Business overdraft is an easy and effective short-term option for many companies to solve cash shortages. It can ensure a company has funds in place and available immediately when something unexpected happens. After an agreement between a company and a bank is made, overdraft occurs when money is withdrawn from a bank account when the available balance goes below zero. The company pays interest for the negative balance at an agreed rate.  A survey \citep{Survey2014} about SMEs in the 28 countries of the European Union showed that SMEs preferred to use bank overdraft, bank loan and trade credit. A report by \cite{IpsosMORI2017} based on the surveys of over 1000 SMEs from 2014 to 2016 in UK revealed that business overdraft accounted for about 20\% --- and ranked second --- of all external finances in the three years.

Given the above account, it is important for a manager to take capital flow and external financing into account when making operational decisions.  In this respect, we contribute to the literature on stochastic lot sizing as follows.
\begin{itemize}
\item We incorporate capital flow and one kind of external financing, i.e. overdraft, in the stochastic lot sizing problem and formulate a stochastic dynamic programming model to obtain optimal solutions.
\item We discuss four inventory controlling policies for this problem and use simulation-genetic algorithm to obtain approximate values of the controlling parameters. 
\item We introduce a simulation-optimization heuristic inspired by the approach originally introduced by \cite{askin1981}.
\item We conduct a comprehensive numerical study to compare stochastic dynamic programming, simulation-genetic algorithm and simulation-optimization heuristic.  
\end{itemize}

The rest of this work is structured as follows. Section 2 reviews the related literature. Section 3 describes the problem setting. Section 4 provides the stochastic dynamic programming model. Section 5 gives the simulation-genetic algorithm approach and Section 6 presents the simulation-optimization heuristic. Section 7 adopts some numerical examples to show the influence of capital flow and overdraft to optimal policy structure. 
A computational study and its results are detailed in Section 8. Finally, Section 9 draws conclusions and outlines future research directions.

\section{Literature review}

Due to its practical relevance, a large body of literature has emerged on the lot sizing problem over the last few decades. After the pioneering work by \cite{wagner1958dynamic} for the single item lot sizing problem with deterministic demands, many extensions were investigated for lot sizing problems, such as multi item settings, capacity constraints, and stochastic demand. Here we limit our discussion to key studies that are relevant in the context of our discussion. 

The stochastic lot sizing problem has been thoroughly investigated in the inventory control literature \cite{Axsater2007}. Under mild assumptions, \cite{scarf1959} proved the optimal policy takes the well-known ($s,S$) form, in which $s$ denotes the reorder point and $S$ is the order-up-to-level. \cite{silver1978inventory} and \cite{askin1981} proposed simple heuristics based
on the least period cost method for the problem under penalty cost, which are stochastic extensions of the well-known Silver–-Meal's heuristic \citep{silver1973heuristic}. \cite{bookbinder1988strategies} discussed three control strategies: the ``static uncertainty'' strategy, in which lot sizing decisions including review intervals and order quantities for each period must be made at the beginning of the first period; the ``static-dynamic uncertainty" strategy, in which review intervals are fixed at the beginning but order quantities for each period are not determined until this period comes; and the ``dynamic uncertainty" strategy, which allows the retailer to decide dynamically at each period whether or not to place an order and how much to order.
%

Many subsequent studies focused on modeling and solving the stochastic lot sizing problem under the three strategies introduced in \cite{bookbinder1988strategies}. In the context of the ``static uncertainty'' strategy, \cite{guan2006branch} 
developed a branch-and-cut algorithm; \cite{tempelmeier2011dynamic} presented modifications of several well-known dynamic lot sizing heuristics for solving the single item stochastic problem;  \cite{koca2015stochastic} used second order cone programming to obtain solutions for a capacitated single item lot sizing problem with controllable processing time. In the context of the ``static-dynamic uncertainty'' strategy, \cite{tarim2004stochastic} proposed a mathematical programming approach to compute near-optimal ``static-dynamic uncertainty'' policy parameters under service level constraints; \cite{tarim2006modelling} extended the previous study to a penalty cost setting, by using a piecewise linear approximation of the cost function; \cite{tempelmeier2007stochastic} proposed a $\beta$ service level constraint and a new computation of holding costs for the problem; \cite{ozen2012static} considered the problem with dynamic fixed-ordering, holding costs and dynamic penalty costs and proved that the optimal policy is base-stock policy; \cite{rossi2015piecewise} applied a new piecewise approximation method in \cite{rossi2014piecewise} to handle a range of service level measures as well as lost sales. The ``dynamic uncertainty" strategy was initially investigated in \cite{askin1981}, who developed an early heuristic; \cite{federgruenandzipkin1984} addressed the stationary demand case, and \cite{bm1999} extended the previous study to handle the non-stationary demand case.

Relevant works taking capital flow or financing into account in inventory management problems are the following. \cite{buzacott2004inventory} analyzed the importance of joint consideration of production and financial decisions via a news vendor model. \cite{chao2008dynamic} investigated a multi period news vendor problem constrained by cash flow and proved the optimal policy is a base stock policy. \cite{gong2014dynamic} extended the model by incorporating short term financing.  \cite{zeballos2013single} built a periodic review inventory problem with working capital constraints, payment delay and multiple sources of financing. \cite{wuttke2016supply} considered the supplier's supply chain finance adoption decisions within a diffusion model and obtained some insights regarding a buyer's optimal decisions regarding timing and payment terms. The above mentioned works are not for lot sizing problem and there are no fixed ordering costs in these models. Considering capital flow constraints, \cite{Chen161} formulated a single item model for deterministic demands with trade credit and devised a dynamic algorithm with heuristic adjustments to solve it.

Our literature survey reveals that no work has so far investigated the stochastic lot sizing problem under non-stationary stochastic demand, capital flow, and external financing. This, together with our initial discussion on the relevance of this topic, motivates our study. 


\section{Problem description}
For convenience, the notations adopted in this paper are listed in Table \ref{table:notations}. Relevant notations will be introduced when needed. 

In our problem, demand is stochastic and non-stationary. For each period $t$, its demand is represented by $D_{t}$, which is a non-negative random variable with probability density function $f_{t}$, cumulative distribution function $F_{t}$, mean $\mu_{t}$, variance $\sigma_{t}^{2}$. Random demand is assumed to be independent over the periods. Unmet demand in any given period is back ordered and satisfied as soon as the replenishment arrives. Excess stock is transferred to next period as inventory and the sell back of excess stock is not allowed.

We assume initial capital quantity of the retailer is $B_{0}$; order delivery lead time is zero; selling price of the product is $p$ and the retailer receives payments only when the products are delivered to its customers. A fixed cost $a$ is charged when placing orders, regardless of the order quantity, and $R_{t}$ is a 0-1 variable to determine whether the retailer makes order at period $t$; a variable cost $v$ is charged on every order unit. End-of-period inventory level for period $t$ is $I_{t}$, and we set $I_{t}^{+}$ to represent $\max\{I_{t},0\}$ and $I_{t}^{-}$ to represent $\max\{-I_{t},0\}$. A variable inventory holding cost $h$ is charged on every product unit carried from one period to the next; per unit stock-out penalty cost is $\pi$; at the beginning of each period $t$, its present capital is $B_{t-1}$, if its initial capital is below zero, the retailer has to pay interests with a rate of $b$. 

\begin{figure}
\centering\footnotesize
\begin{overpic}[scale=0.8]{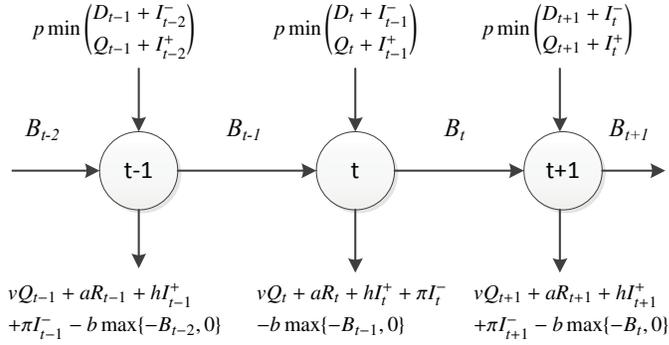}
\put(10,41){$ p\min\dbinom{D_{t-1}+I_{t-2}^{-}}{Q_{t-1}+I_{t-2}^{+}}$}
\put(40,41){$p\min\dbinom{D_{t}+I_{t-1}^{-}}{Q_{t}+I_{t-1}^{+}}$}
\put(66,41){$p\min\dbinom{D_{t+1}+I_{t}^{-}}{Q_{t+1}+I_{t}^{+}}$}
\put(7,8){$vQ_{t-1}+a R_{t-1}+h I_{t-1}^{+}$}
\put(7,4){$+\pi I_{t-1}^{-}-b\max\{-B_{t-2},0\}$}
\put(38,8){$vQ_{t}+a R_{t}+h I_{t}^{+}+\pi I_{t}^{-}$}
\put(38,4){$-b\max\{-B_{t-1},0\}$}
\put(65,8){$vQ_{t+1}+a R_{t+1}+h I_{t+1}^{+}$}
\put(65,4){$+\pi I_{t+1}^{-}-b\max\{-B_{t},0\}$}
\end{overpic}
\caption{Inflows and outflows of capital at periods $t-1$, $t$, $t+1$}\label{fig:capitalflow}
\end{figure}

The actual sales quantity in period $t$ is $\min\big\{D_{t}+I_{t-1}^{-},Q_{t}+I_{t-1}^{+}\big\}$, where $D_{t}+I_{t-1}^{-}$ is demand plus backorder in period $t$ and $Q_{t}+I_{t-1}^{+}$ is the total available stock in period $t$. End-of-period capital $B_{t}$ for period $t$ is defined as its initial capital $B_{t-1}$, plus payments by customers for the realized demand in this period $p\min\big\{D_{t}+I_{t-1}^{-},Q_{t}+I_{t-1}^{+}\big\}$, minus the payments to suppliers and this period's fixed ordering cost, holding and backorder costs $vQ_{t}+a R_{t}+h I_{t}^{+}+\pi I_{t}^{-}$, and minus the interest paid if its initial capital is negative $b\max\{-B_{t-1},0\}$. Full expression of $B_{t}$ is given by Eq. (1), and the inflows and outflows of capital from period $t-1$ to period $t+1$ is detailed shown by Figure \ref{fig:capitalflow}.
\begin{equation}
B_{t}=B_{t-1}+p\min\big\{D_{t}+I_{t-1}^{-},Q_{t}+I_{t-1}^{+}\big\}-\left(vQ_{t}+a R_{t}+h I_{t}^{+}+\pi I_{t}^{-}\right)-b\max\{-B_{t-1},0\},\label{eq:capitalflow}
\end{equation}

As for the final capital in the whole planning horizon, we defined it as the end-of period capital $B_{T}$, minus the interest paid if $B_{T}$ is negative, which is:
\begin{equation}
B_{T+1}=B_{T}-b\max\{-B_{T},0\}.\label{eq:finalcapital}
\end{equation}

Our aim is to find a replenishment plan that maximizes the expected final capital increment, i.e. $E(B_{T+1})-B_0$. 

\begin{table}[!ht]
\tbl{Notations adopted in our paper.}
{\begin{tabular}[b]{p{3cm}<{\raggedright}p{7cm}<{\raggedright}}
\toprule
Notations  &Description\\
\colrule
Indices  &\\
$t$        &Period index, $t=1,2,\dots,T$\\
\specialrule{0em}{3pt}{3pt}
Problem parameters  &\\
\specialrule{0em}{1pt}{1pt}
$B_{0}$  &Initial capital\\
$I_{0}$  &Initial inventory, we assume $I_{0}=0$\\
$I_{t}^{+}$  &$\max\{I_{t},0\}$\\
$I_{t}^{-}$  &$\max\{-I_{t},0\}$\\
$p$      &Product selling price\\
$a$      &Fixed ordering cost\\
$v$      &Unit variable ordering cost\\
$h$      &Unit inventory holding cost\\
$\pi$    &Unit penalty cost for back orders\\
$b$      &Interest rate for negative initial capital\\
\specialrule{0em}{3pt}{3pt}
Random variables  &\\
\specialrule{0em}{1pt}{1pt}
\multirow{3}*{$D_{t}$}      &Random demand at period $t$ with probability density function $f_{t}(D_{t})$,cumulative distribution function $F_{t}(D_{t})$, mean $\mu_{t}$, variance $\sigma_{t}^{2}$\\
\specialrule{0em}{3pt}{3pt}
State variables  &\\
\specialrule{0em}{1pt}{1pt}
$I_{t}$      &End-of-period inventory for period $t$\\
$B_{t}$      &End-of-period capital for period $t$\\
\specialrule{0em}{3pt}{3pt}
Decision variables  &\\
\specialrule{0em}{1pt}{1pt}
$Q_{t}$ &Order quantity at the beginning of period $t$\\
$R_{t}$ &whether the retailer orders at period $t$\\
\multirow{2}*{$S_{t}$} &Order up to level at the beginning of period $t$, and $S_{t}=I_{t-1}+Q_{t}$\\
$s_{t}$  &Threshold of the inventory level for $(s, S)$ policy\\
$\overline{Q}_{t}$  &Maximum order quantity for $(s, \overline{Q}, S)$ policy\\
\botrule
\end{tabular}}\label{table:notations}
\end{table}

\section{A stochastic dynamic programming approach}
In this section, we apply stochastic dynamic programming (SDP) to solve the problem.

{\bf States.}
The system state at the beginning of period $t$ is represented by initial inventory $I_{t-1}$ and initial capital quantity $B_{t-1}$. Note that $I_{t-1}$ and $B_{t-1}$ can be negative.

{\bf Actions.}
The action at period $t$ is the production quantity $Q_{t}$, given initial inventory $I_{t-1}$ and initial capital quantity $B_{t-1}$. 

{\bf State transition function.}
The inventory and capital at the end of period $t$ is determined by initial inventory $I_{t-1}$, initial capital $B_{t-1}$, demand $D_t$ and action $Q_{t}$. The state transition of the system for $B_{t}$ and $I_{t}$ is described by the following equations.
\begin{align}
B_{t}=&B_{t-1}+p\min\big\{D_{t}+I_{t-1}^{-},Q_{t}+I_{t-1}^{+}\big\}-\left(a\delta_{t}+hI_{t}^{+}+\pi I_{t}^{-}+vQ_{t}\right)-b\max\{-B_{t-1},0\},\quad\\
I_{t}=&I_{t-1}+Q_{t}-D_{t},\quad\label{eq:Iflow}
\end{align}

Capital transition equation is same as the expression of end-of-period capital $B_{t}$. 

{\bf Immediate profit.}
The immediate profit for period $t$ is the expected capital increment during this period. Given $B_{t-1}$, $I_{t-1}$ and $Q_{t}$, immediate profit $\Delta B_{t}$ can be expressed as
\begin{align}
\Delta B_{t}^{Q_{t}}(I_{t-1},B_{t-1})=E(B_{t})-B_{t-1},
\end{align}

{\bf Functional Equation.}
Define $F_{t}(B_{t-1},I_{t-1})$ as the maximum expected capital quantity increment during periods $t,t+1,\dots,T$, when the initial inventory and capital of period $t$ are $I_{t-1}$ and $B_{t-1}$, respectively. The functional equation is expressed as:
\begin{align}
F_{t}(B_{t-1},I_{t-1})=\max\limits_{Q_{t}\geq 0}\left\{\Delta B_{t}+F_{t+1}(B_{t},I_{t})\right\},\quad t=1,2,\dots,T,\label{eq:functionalequation}
\end{align}

By the definition of final capital in Eq. \eqref{eq:finalcapital}, the boundary condition for the functional equation is:
\begin{equation}
F_{T+1}(B_{T},I_{T})=-b\max\{-B_{T},0\}.
\end{equation}
Our aim is to maximise the expected capital increment $F_{1}(B_0,I_0)$ over the planning horizon given initial capital $B_0$ and inventory $I_0$.
\subsection{A numerical example}
Here we give a numerical example to show the results of SDP. Values of parameters are listed in Table \ref{table:parametervalues1}. For convenience, there are 3 periods in total, and two levels of demand in each period with equal probability 0.5. Details of demand levels are presented in Table \ref{table:demands}.

\begin{table}[ht]
\tbl{Parameter values.}
{\begin{tabular}{*{8}{m{0.8cm}}}
\toprule
$B_{0}$&$I_{0}$ & $p$ & $a$ & $v$ & $h$ &$\pi$&b\\
\colrule
 5  &0       & 5      & 10       & 1      & 1  &2 &20\%\\
\botrule
\end{tabular}}\label{table:parametervalues1}
\end{table}

\begin{table}[ht]
\tbl{Demand values.}
{\begin{tabular}{*{4}{m{1.2cm}}}
\toprule
&$D_1$ &$D_2$ &$D_3$  \\
\colrule
level 1  &1       & 1      & 1\\
level 2  &2       & 2      & 2\\
\botrule
\end{tabular}}\label{table:demands}
\end{table}

From SDP, optimal expected final capital increment is 1.75. For 5 random demand realization samples, results of SDP are shown by Table \ref{tab:solutionSDPExamplereSample}. 
\begin{table}[ht]
\tbl{Solutions of SDP for 5 demand realization samples.}
{\begin{tabular}{lrrrrrrc}
\toprule
&$D_1$ &$D_2$ &$D_3$ &$Q_{1}$  &$Q_2$ &$Q_3$ &$E(B_{T+1})-B_0$\\
\colrule
sample 1  &2 & 1 & 2  &0  &5    &0  &3.8  \\
sample 2  &2 & 1  &1  &0  &5  &0    &-2.2 \\
sample 3  &2 & 2  &2  &0  &5   &0   &3.0 \\
sample 4  &1 & 1  &2  &0 &4   &0    &1.4\\
sample 5  &1 & 2  &1  &0  &4  &0    &3.0\\
\botrule
\end{tabular}}\label{tab:solutionSDPExamplereSample}
\end{table}

\section{A simulation-based genetic algorithm approach}

Because of states explosion, it is time consuming for stochastic dynamic programming to obtain a final result. Therefore, we propose four inventory controlling policies based on three strategies of stochastic lot sizing problem (\cite{bookbinder1988strategies}) to solve our problem.

\begin{enumerate}
\item {\bf $(R, Q)$ policy.} This is a ``static uncertainty" strategy, in which the review intervals and ordering quantities for replenishment periods are fixed at the beginning of planning horizon. For this policy, the decision variables are review intervals $R_{t}$
and ordering quantities $Q_{t}$.
\item {\bf $(R, S)$ policy.} This is a ``static-dynamic" strategy, in which the review intervals are fixed at the beginning of the planning horizon, but the order quantities are not fixed. Order quantity  $Q_{t}$ is related with the inventory level at the review intervals $I_{t-1}$ and the fixed order-up-to levels $S_{t}$, which is $\max\{0,S_t-I_{t-1}\}$.
\item {\bf $(s, S)$ policy.} This is a ``dynamic uncertainty" strategy policy: $(s, S)$ policy, in which the review intervals and ordering  quantity are neither fixed. If initial inventory level at period $t$ is above $s_t$, order nothing in this period; else order to the order-up-to level $S_t$. 

\item {\bf $(s, \overline{Q}, S)$ policy.} This is a variant of the previous ``dynamic uncertainty" strategy, in which we impose an upper bound $\overline{Q}$ for the maximum ordering quantity. 
\end{enumerate}
 
Details of order quantity $Q_t$ in $(s, S)$ policy is presented by the formula below.
 \begin{align}
 Q_{t}=\begin{cases}
 0\quad &I_{t-1}\geq s_t\\
 \max\Big\{0,{S_t-I_{t-1}}\Big\}\quad &I_{t-1}<s_t.
 \end{cases}
 \end{align}

For the $(s, \overline{Q}, S)$ policy, order quantity for each period $t$ is determined by Eq. \eqref{eq:sQS}.
 \begin{align}
 Q_{t}=\begin{cases}
 0\quad &I_{t-1}\geq s_t\\
 \max\Big\{\overline{Q}_t,\max\{0,S_t-I_{t-1}\}\Big\}\quad &I_{t-1}<s_t.
 \end{cases}\label{eq:sQS}
 \end{align}
 
Since our problem explicitly models capital flows, one may be inclined to adopt an inventory policy in which decisions are dependent on available stock and capital at the beginning of each period. However, because end-of-period capital of each period is, in fact, solely determined by its initial capital, order quantity, and previous demand realizations, as we will show, a control policy that only takes into account initial stock level in each period --- essentially an ($s,S$) policy --- and that sets an upper bound to the maximum possible replenishment quantity is able to achieve near-optimal performances. Furthermore, even in absence of an upper bound to the maximum possible replenishment quantity, an ($s,S$) policy appears to perform well.

\subsection{Genetic algorithm settings}
It is difficult to obtain values of controlling parameters for these policies from mathematical optimization directly. Therefore, we solve this problem by Monte-Caro simulation and genetic algorithm to obtain near-optimal parameter values. Informally speaking, our approach proceeds as follows. We generate a sufficiently large random demand samples and obtain values of controlling parameters by maximizing the average final capital for these samples via genetic algorithm; then we simulate the controlling parameters over 100,000 benchmark random demands to estimate final expected capital increment.  

Genetic algorithm (GA) is a global search procedure that searches from one population of solutions to another, in which good solutions are selected to evolve to the next generation while some mutations take place to avoid falling into local optimum traps \citep{dorsey1995genetic}. For our problem, besides the lower and upper bounds of the controlling parameters, there are no extra constraints. It is not difficult for genetic algorithm to generate feasible populations and find a good solution. Settings of genetic algorithm we adopted are listed below; note that to assess the validity of these settings we will carry out stability analysis \citep{kaut2007stability} in our computational study.
\begin{enumerate}
\item {\bf Fitness function, bounds of variables.} The fitness function is the maximum average final capital increment of the samples. $R_{t}$ is a 0-1 variable; lower bound for $Q_{t}$ and $\overline{Q}_t$ are 0; lower bounds for $s_{t}$ and ${S}_{t}$ are negative enough numbers; upper bounds for $Q_{t}$, $\overline{Q}_t$, $s_{t}$, and ${S}_{t}$ are large enough numbers.

\item {\bf Population.} Population type is double vector real coding and population size is 200. Initial populations are randomly generated between the lower and upper bounds of each controlling parameter. 

\item {\bf Fitness scaling.} Fitness scaling is based on the rank of fitness scores of each individual population, which can remove the effect of the spread of the raw scores. The rank of an individual is its position in the sorted scores. An individual with rank $r$ has scaled score $1/\sqrt{r}$. This scaling method makes poorly ranked individuals more nearly equal in scores.

\item {\bf Selection.} We use roulette selection method to chooses parents for the next generation based on their scaled fitness values. 

\item {\bf Reproduction and crossover.} The minimum number of elite individuals that are guaranteed to survive to the next generation is 10. Other children are produced by crossover of the selected parents. The Crossover rate is 0.8. Crossover method is scattered, which first creates a random binary vector and then selects the genes where the vector is a 1 from the first parent, and the genes where the vector is a 0 from the second parent, and combines the genes to form a child.

\item {\bf Mutation.} The mutation method is Gaussian, which add a random number to the selected parents. The random number is generated from a Gaussian distribution with mean 0 and shrinking standard deviation. Initial standard deviation $\delta_{0}$ is 10 and the shrinking of standard deviation for the $k_{th}$ generation is shown by the following equation, in which \emph{generations} is the maximum generations initially set for the genetic algorithm to perform.
\begin{equation}
\delta_{k}=\delta_{k-1}\Big(1-\frac{k}{generations}\Big).
\end{equation}

\item {\bf Migration.} Migration takes place every 20 generations forward, and the best 20\% individuals from one subpopulation replace the worst individuals in another subpopulation.

\item {\bf Stopping criteria.} Maximum generations is 10000 and function tolerance is 0.000001. If the average change in the fitness function value over 50 generations is less than function tolerance, the algorithm stops.
\end{enumerate}

\subsection{A numerical example}

For the same numerical example solved by SDP, solutions of genetic algorithm for the four controlling policies are given by Table \ref{tab:solutionGAexample}. 
\begin{table}[ht]
\tbl{Solutions of genetic algorithm.}
{\begin{tabular}{lllllllc}
\toprule
& & &&&& &$E(B_{T+1})-B_0$ \\
\colrule
\multirow{2}*{policy $(s,\overline{Q},S)$}  &$s_1=-1$ & $s_2=0$  & $s_3=4$ &$S_1=7$ & $S_2=3$  & $S_3=0$ &\multirow{2}*{1.29}\\
&$\overline{Q}_1=9$&$\overline{Q}_2=7$&$\overline{Q}_3=8$&&&&\\
\specialrule{0em}{1pt}{1pt}
policy $(s,S)$  &$s_1=0$ & $s_2=7$  & $s_3=0$ &$S_1=5$ & $S_2=3$  & $S_3=3$ &1.29\\
policy $(R,S)$  &$R_1=0$ & $R_2=1$  &$R_3=0$   &$S_1=0$ & $S_2=3$  & $S_3=0$ &1.29\\
policy $(R,Q)$  &$R_1=0$ & $R_2=1$  &$R_3=0$   &$Q_1=0$ & $Q_2=5$  & $Q_3=0$ &0.65\\
\botrule
\end{tabular}}\label{tab:solutionGAexample}
\end{table}

For the same 5 random demand realization samples, final capital increments of four controlling policies are provided by Table \ref{tab:solutionGAExamplereSample}. 
\begin{table}[ht]
\tbl{Expected final capital increment for 5 demand realization samples.}
{\begin{tabular}{lrrrrrrr}
\toprule
&$D_1$ &$D_2$ &$D_3$ &$(s,\overline{Q},S)$& $(s,S)$ & $(R,S)$ & $(R,Q)$ \\
\colrule
sample 1  &2 & 1 & 2 &3.8 &3.8  &3.8    &3.8    \\
sample 2  &2 & 1  &1 &-2.2 &-2.2  &-2.2   &-2.2     \\
sample 3  &2 & 2  &2 &3.0 &3.0  &3.0   &3.0    \\
sample 4  &1 & 1  &2 &1.4 &1.4  &1.4   &-2.0    \\
sample 5  &1 & 2  &1 &3.0 &3.0  &3.0  &0.0    \\
\botrule
\end{tabular}}\label{tab:solutionGAExamplereSample}
\end{table}
It shows, policy $(s,\overline{Q},S)$, policy $(s,S)$ and $(R,S)$ obtain the same final expected capital increment as SDP, while policy $(R, Q)$ performs a little worse.

\section{A simulation-optimization heuristic approach}
In this section, we present a simulation-optimization heuristic for our problem that is based on Silver's heuristic \citep{silver1978inventory}. To describe the simulation-optimization heuristic, we first define the concept of \emph{replenishment cycle}.
\newtheorem{definition}{Definition}
\begin{definition}
A replenishment cycle $(t,r)$ is the time interval between two consecutive replenishments executed at period $t$ and at period $r + 1$; the last replenishment cycle is $(t,T)$.
\end{definition}

The idea behind Silver's lot sizing heuristic is to view the average cost per period as a function of the length of a replenishment cycle. Assuming this function is convex with respect to the length of the replenishment cycle, it is possible to determine the optimal length of the next replenishment cycle. With regard to our profit-maximization problem, we view the average capital increment per period is concave to the replenishment cycle length. For the computation of capital increment in a replenishment cycle, we have the following property.

\newtheorem{lemma}{Lemma}
\begin{lemma}
The expected total capital increment during replenishment cycle $(t,r)$ given initial inventory $I_{t-1}$ and initial capital $B_{t-1}$, is concave in ordering quantity $Q_{t}$ for $Q_{t}>0$.
\end{lemma}\label{lemma:concave}

\begin{proof}
Total expected capital increment in replenishment cycle $(t,r)$ is
\begin{equation}
\Delta B^{Q_{t}}_{t,r}(I_{t-1},B_{t-1})=\Delta B_{t}^{Q_{t}}(I_{t-1},B_{t-1})+\Delta B_{t+1}^{0}(I_{t},B_{t})+\dots+\Delta B_{r}^{0}(I_{r-1},B_{r-1}),
\end{equation}

For period $t$,
\begin{equation}
\Delta B_{t}^{Q_{t}}(I_{t-1},B_{t-1})=E\left(p\min\big\{D_{t}+I_{t-1}^{-},Q_{t}+I_{t-1}^{+}\big\}-\left(a+hI_{t}^{+}+\pi I_{t}^{-}+vQ_{t}\right)-b\max\{-B_{t-1},0\}\right),
\end{equation}

Given $I_{t-1}$ and $B_{t-1}$, it can be easily proved that the above formula is concave in $Q_{t}$. For period $n\in\{t+1,t+2,\dots,r\}$,
\begin{equation}
\Delta B_{n}^{0}(I_{n-1},B_{n-1})=E\left(p\min\big\{D_{n}+I_{n-1}^{-},I_{n-1}^{+}\big\}-\left(hI_{n}^{+}+\pi I_{n}^{-}\right)-b\max\{-B_{n-1},0\}\right).
\end{equation}

The above function is separable, and its separable items are all concave piecewise linear functions
of $Q_t$: the first term $p\min\big\{D_{n}+I_{n-1}^{-},I_{n-1}^{+}\big\}$ is concave in $Q_{t}$, the second term $-(hI_{n}^{+}+\pi I_{n}^{-})$ is concave in $Q_{t}$, and the third term $-b\max\{-B_{n-1},0\}$ is also concave because $B_{n-1}$ is concave. This function is
therefore concave in $Q_t$.
\end{proof}

For the simulation-optimization heuristic, we use Monte-Carlo simulation to approximate $E(I_t)$ in a replenishment
cycle by randomly generating an appropriate number of demand samples. Using this sample-based approximation we can estimate expected total capital increment of the replenishment cycle and therefore
develop a simulation-based extension of Silver’s heuristic. We apply line search to compute the optimal order quantity $Q_{t}$ for replenishment
cycle $(t,r)$ as well as the associated expected total capital increment based on Lemma \ref{lemma:concave}. We increase the value of $r$, starting from $t$, until the expected average capital increment per period first decreases. A pseudo-code of the algorithm is given in Algorithm \ref{algorithm}. Computational complexity of our simulation-optimization heuristic is comparable to that discussed for the approach in \citep{gutierrez2017simple}, which is $O(NTlog(\overline{Q}/\lambda))$, where $N$ is the number of samples used, $T$ is the length of planning horizon, $\overline{Q}$ is the upper bound of ordering quantity and $\lambda$ is the step length used in the line search.

\SetKwRepeat{Do}{do}{while}
\begin{algorithm}
\caption{Simulation-optimization heuristic}\label{algorithm}
\KwData{current period $t$, initial inventory $I_{t-1}$, initial capital $B_{t-1}$}
\KwResult{optimal order quantity $Q^{\ast}_{t}$}
$r\leftarrow t$\;
$\Delta B^{\ast}\leftarrow -\infty$\;
\Do{$\Delta B\geq \Delta B^{\ast}$ and $r\leq T$}{
$Q\leftarrow\arg\max_{Q\geq 0}\Delta B^{Q}_{t,r}(I_{t-1},B_{t-1})$\;
$\Delta B\leftarrow \Delta B^{Q}_{t,r}(I_{t-1},B_{t-1})/(r-t+1)$\;
\If{$\Delta B\geq \Delta B^{\ast}$}
{$Q^{\ast}\leftarrow Q$\;
$\Delta B^{\ast}\leftarrow \Delta B$\;}
$r\leftarrow r+1$\;
}
\end{algorithm}


\subsection{A numerical example}

In the same numerical example as SDP and genetic algorithm, expected final capital increment by simulation-optimization heuristic is -1.25. Results of simulation-optimization heuristic for the same 5 random demand realization samples are provided by Table \ref{tab:solutionSimExamplereSample}.

\begin{table}[ht]
\tbl{Solutions of simulation-optimization heuristic for 5 demand realization samples.}
{\begin{tabular}{lrrrrrrr}
\toprule
&$D_1$ &$D_2$ &$D_3$ &$Q_{1}$  &$Q_2$ &$Q_3$ &$E(B_{T+1})-B_0$\\
\colrule
sample 1  &2 & 1 & 2  &0  &4    &0  &-1.0  \\
sample 2  &2 & 1  &1  &0  &4  &0    &1.0 \\
sample 3  &2 & 2  &2  &0  &4   &0   &-2.0 \\
sample 4  &1 & 1  &2  &0 &3   &0    &-3.2\\
sample 5  &1 & 2  &1  &0  &3  &0    &-2.0\\
\botrule
\end{tabular}}\label{tab:solutionSimExamplereSample}
\end{table}

Compared with SDP and four controlling policies by genetic algorithm, simulation-optimization heuristic performs a little worse for this numerical example.

\section{Numerical examples}

In this section, we illustrate the importance of modeling capital flows and business overdraft. By using a selection of instances, we demonstrate the impact of changes in problem parameters on the structure of the control action. 

We consider 6 periods, demands in each period are independent and follow Poisson distribution. Mean values of the demands of each period are listed in Table \ref{table:meanand deviation demands}, other relevant parameter values are given by Table \ref{table:parametervalues}.

\begin{table}[ht]
\tbl{Mean values of demands.}
{\begin{tabular}{*{7}{m{0.8cm}}}
\toprule
period &1 & 2 & 3 & 4 & 5 &6\\
\colrule
$\tilde{D}_{t}$        & 3    & 4      & 3      & 5    & 4  &3 \\
\botrule
\end{tabular}}\label{table:meanand deviation demands}
\end{table} 
\begin{table}[ht]
\tbl{Parameter values.}
{\begin{tabular}{*{7}{m{0.8cm}}}
\toprule
$I_0$&$p$ & $a$ & $v$ & $h$ &$\pi$ & b\\
\colrule
0& 4      & 12       & 2      & 1  &3 &20\%\\
\botrule
\end{tabular}}\label{table:parametervalues}
\end{table}

If $I_0=0$, $B_0=0$, optimal final capital increment is -3.98. If $I_0=0$, $B_0=20$, optimal final capital increment is 5.52. If demand realizations are same as expected demands, optimal policy for the two case with different initial capital is presented by Table \ref{table:policyforcapital}.

\begin{table}[ht]
\tbl{Optimal solutions for two different initial capital.}
{\begin{tabular}{lrrrrrrr}
\toprule
 &$t=1$ & $t=2$ & $t=3$ & $t=4$ & $t=5$ &$t=6$&$E(B_{T+1})-B_0$\\
\colrule
$Q_{t}$ ($B_0=0$)        & 0    & 9      & 0      & 11    & 0  &0 &-3.98\\
$Q_{t}$ ($B_0=20$)        & 9    & 0      & 0      & 12    & 0  &0 &5.52\\
\botrule
\end{tabular}}\label{table:policyforcapital}
\end{table} 

Table \ref{table:policyforcapital} shows available capital may have substantial impact on the structure of the optimal policy and maximum final capital increment. If $I_0=0$, $B_0=0$, but interest rate $b=5\%$, optimal final capital increment is 4.90. With same demands realizations, solutions of the two cases with different interest rates are given in Table \ref{table:policyforinterest}.

\begin{table}[ht]
\tbl{Optimal solutions for two different interest rates.}
{\begin{tabular}{lrrrrrrr}
\toprule
 &$t=1$ & $t=2$ & $t=3$ & $t=4$ & $t=5$ &$t=6$&$E(B_{T+1})-B_0$\\
\colrule
$Q_{t}$ ($b=5\%$)        & 0    & 9      & 0      & 11    & 0  &0 &4.90\\
$Q_{t}$ ($b=20\%$)        & 8    & 0      & 0      & 13    & 0  &0 &-3.98\\
\botrule
\end{tabular}}\label{table:policyforinterest}
\end{table} 

Results in Table \ref{table:policyforinterest} show interest rate can also substantially affect the structure of optimal policy. Other combinations of parameters, e.g. ``contribution margin'' that is selling price minus unit variable production cost, can also influence optimal policy. With $I_0=0$, $B_0=0$, $b=20\%$, optimal solutions for two different margins are presented by Table \ref{table:policyformargins} (for margin = 3, $v=2$, $p=5$; for margin = 5, $v=1$, $p=6$).

\begin{table}[ht]
\tbl{Optimal solutions for two different margins.}
{\begin{tabular}{lrrrrrrr}
\toprule
 &$t=1$ & $t=2$ & $t=3$ & $t=4$ & $t=5$ &$t=6$&$E(B_{T+1})-B_0$\\
\colrule
$Q_{t}$ (margin=3)        & 0    & 10      & 0      & 9    & 0  &0 &-7.94\\
$Q_{t}$ (margin=5)        & 8    & 0      & 0      & 13    & 0  &0 &44.78\\
\botrule
\end{tabular}}\label{table:policyformargins}
\end{table} 




The above results reveal that capital availability and interest rates impact retailer's optimal policy structure and expected final capital increment in the stochastic lot sizing problem.

\section{Computational study}

In this section we present an extensive computational study to investigate the effectiveness of our approaches.


\subsection{Test bed}
\begin{figure}[!ht]
\begin{center}
\includegraphics[width=9cm,height=14cm]{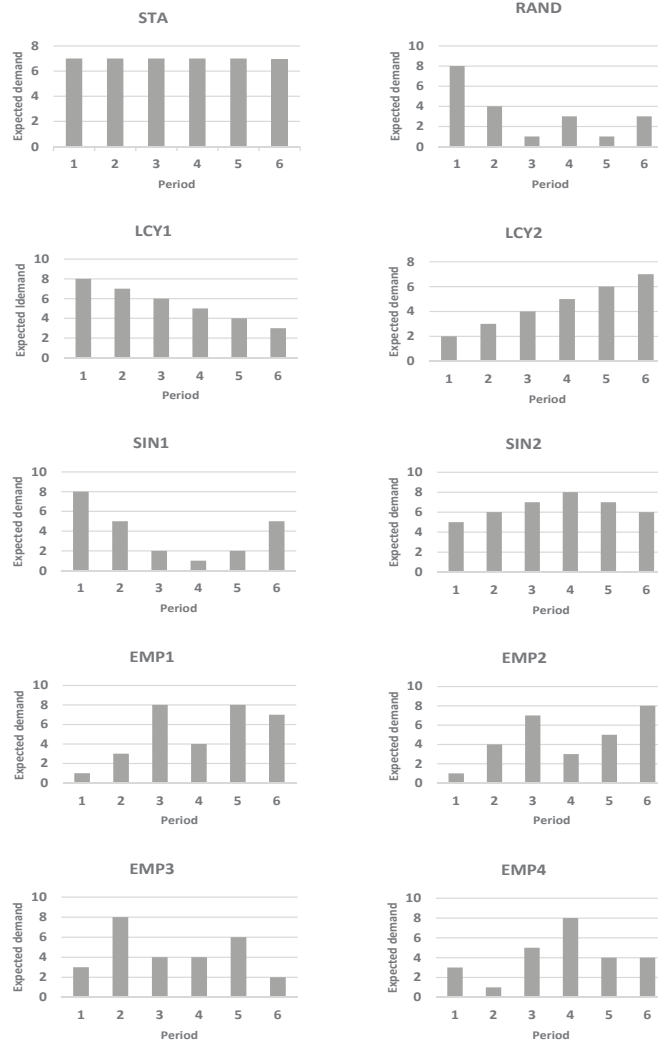}
\caption{Demand patterns in our computational analysis.}\label{fig:expecteddemand}
\end{center}
\end{figure}

The test beds are adopted from \cite{rossi2015piecewise}. There are 10 demand patterns for numerical analysis: 2 life cycle patterns (LCY1 and LCY2), 2 sinusoidal patterns (SIN1 and SIN2), 1 stationary pattern (STA), 1 random pattern (RAND), 4 empirical patterns (EMP1, EMP2, EMP3, EMP4). To ensure SDP can solve these instances in reasonable time, we rescale original demands in \cite{rossi2015piecewise} and select 6 successive periods for testing. The values of expected demands for different patterns are given by Table \ref{table:expecteddemand} and Figure \ref{fig:expecteddemand}, respectively.

\begin{table}[!ht]
\tbl{Expected demands of different patterns.}
{\begin{tabular}[b]{*{8}{m{1cm}}}
\toprule
demand pattern &1 & 2 & 3 & 4 & 5 &6 & Total\\
\colrule
STA   & 7   & 7      & 7      & 7    & 7  &7 &42\\
LCY1   & 8    & 7     & 6     & 5   & 4 &3 &33\\
LCY2  & 2    & 3      & 4     &5    & 6 &7&27\\
SIN1  & 8  & 5      & 2      & 1    & 2  &5&23\\
SIN2  & 5   & 6      & 7      & 8    & 7  &6&39\\
RAND  & 8   & 4      & 1      & 3    & 1  &3&20\\
EMP1  & 1   & 3      & 8      & 4    & 8  &7&31\\
EMP2  & 1	&4	     &7	  &3	  & 5	&8 &28\\
EMP3  & 3	&8	     &4	 &4	&6	&2  &27\\
EMP4  & 3	&1     &5	     &8	&4	&4 &25\\
\botrule
\end{tabular}}\label{table:expecteddemand}
\end{table}

In our tests, demands follow Poisson distribution. Two possible levels are set for each parameter as listed by Table \ref{table:parameterValues}, except unit holding cost $h$, which is fixed to be 1. The values of parameters are set to make most numerical instances encounter capital shortage in the first three periods. For each set of those parameters combinations, 10 demand patterns are tested. Therefore, There are  640 numerical cases in total. 

\begin{table}[!ht]
\tbl{Parameter values.}
{\begin{tabular}[b]{*{7}{m{1.2cm}}}
\toprule
&$B_{0}$  & $p$   &$a$       &$v$    &$\pi$   &$b$\\
\colrule
level 1 &0   & 5   & 10      & 1      & 2    & 0.05  \\
level 2 &20   & 10    & 15     & 2     & 4   & 0.2\\
\botrule
\end{tabular}}\label{table:parameterValues}
\end{table}

\subsection{GA Stability test}
We assess stability of solutions obtained by genetic algorithm on random samples by analysing in-sample and out-of-sample stability. In-sample stability means the final capital increment of genetic algorithm should not vary across  demands scenarios of the same size. Out-of-sample stability refers to that when the solutions are simulated on the benchmark demands, the expected final capital increment should be same among all demands scenarios \citep{kaut2007stability}. 

\begin{figure}[!ht]
\centering
\subfigure[Mean STD for different number of scenarios.]{
\begin{tikzpicture}
\pgfplotsset{every axis legend/.append style={
at={(0.5,1.03)},
anchor=south},every axis y label/.append style={at={(0.07,0.5)}}}
\begin{axis}[title=(a) Mean STD for ,xlabel=Num of scenarios,
    ylabel=Mean STD,xtick =data,legend columns=4,legend style={font=\tiny},font=\footnotesize,scale=0.8]
\addplot table [x=a, y=sQS-std-in,, col sep=comma] {mydata.csv};
\addplot table [x=a, y=sS-std-in, col sep=comma] {mydata.csv};
\addplot table [x=a, y=RS-std-in, col sep=comma] {mydata.csv};
\addplot table [x=a, y=RQ-std-in, col sep=comma] {mydata.csv};
\addplot table [x=a, y=sQS-std-out, col sep=comma] {mydata.csv};
\addplot table [x=a, y=sS-std-out, col sep=comma] {mydata.csv};
\addplot table [x=a, y=RS-std-out, col sep=comma] {mydata.csv};
\addplot table [x=a, y=RQ-std-out, col sep=comma] {mydata.csv};
\legend{s$\overline{Q}$S-in, sS-in, RS-in, RQ-in, s$\overline{Q}$S-out, sS-out, RS-out, RQ-out}
\end{axis}
\end{tikzpicture}}
~~~~
\subfigure[Mean RMSE for different number of scenarios.]{
\begin{tikzpicture}
\pgfplotsset{every axis legend/.append style={
at={(0.5,1.03)},
anchor=south},
every axis y label/.append style={at={(0.07,0.5)}}}
\begin{axis}[xlabel=Num of scenarios,
    ylabel=Mean RMSE,xtick =data,legend columns=4,legend style={font=\tiny},font=\footnotesize,scale=0.8]
\addplot table [x=a, y=sQS-rm-in,, col sep=comma] {mydata.csv};
\addplot table [x=a, y=sS-rm-in, col sep=comma] {mydata.csv};
\addplot table [x=a, y=RS-rm-in, col sep=comma] {mydata.csv};
\addplot table [x=a, y=RQ-rm-in, col sep=comma] {mydata.csv};
\addplot table [x=a, y=sQS-rm-out, col sep=comma] {mydata.csv};
\addplot table [x=a, y=sS-rm-out, col sep=comma] {mydata.csv};
\addplot table [x=a, y=RS-rm-out, col sep=comma] {mydata.csv};
\addplot table [x=a, y=RQ-rm-out, col sep=comma] {mydata.csv};
\legend{s$\overline{Q}$S-in, sS-in, RS-in, RQ-in, s$\overline{Q}$S-out, sS-out, RS-out, RQ-out}
\end{axis}
\end{tikzpicture}}
\caption{Stability test results for different number of scenarios.}\label{fig:InsampleOutsample}
\end{figure}
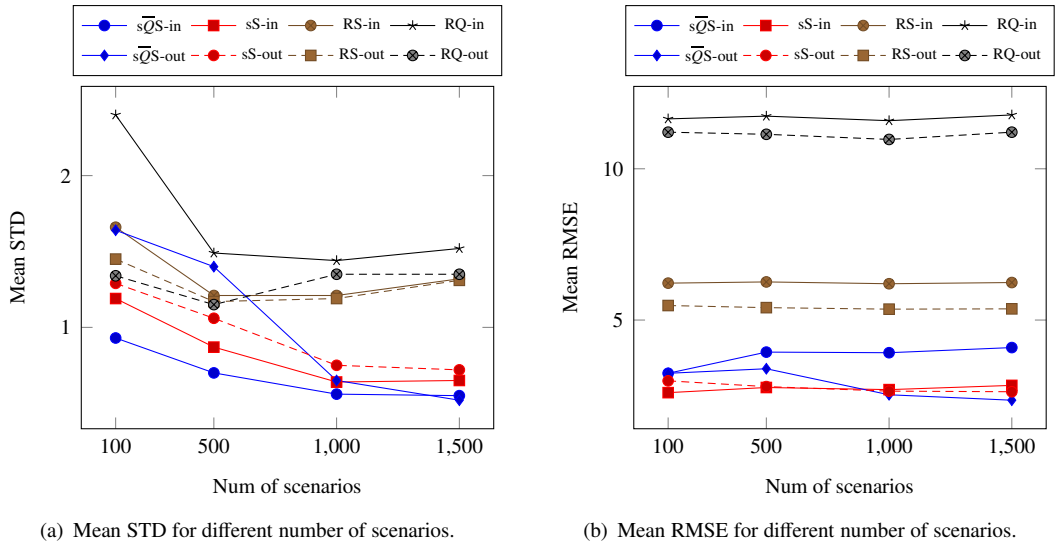

We randomly select 32 cases from the test beds to do stability tests. For a given size of scenarios for each case, we run genetic algorithm for 10 different samples. Like \cite{sexton1998toward}, mean of standard deviation (STD) and mean of root mean square error (RMSE) of the 32 cases are adopted as the stability criteria. Results of stability tests are presented by Table \ref{table:stabilitytest} and Figure \ref{fig:InsampleOutsample}.\footnote{In Figure \ref{fig:InsampleOutsample}, "in" means in-sample, "out" means out-of-sample}

\begin{table}[!ht]
\tbl{In-sample and out-of-sample tests results}
{\begin{tabular}[b]{{m{0.8cm}}*{16}{m{0.5cm}}}
\toprule
\multirow{4}*{scenarios} &\multicolumn{8}{c}{In-sample test} &\multicolumn{8}{c}{Out-of-sample test }\\
\cmidrule(lr){2-9}\cmidrule(lr){10-17}
&\multicolumn{4}{c}{Mean STD}&\multicolumn{4}{c}{Mean RMSE}&\multicolumn{4}{c}{Mean STD}&\multicolumn{4}{c}{Mean RMSE}\\
\cmidrule(lr){2-5}\cmidrule(lr){6-9}\cmidrule(lr){10-13}\cmidrule(lr){14-17}
     &$(s\overline{Q}S)$ &$(sS)$   &$(RS)$       &$(RQ)$ &$(s\overline{Q}S)$ &$(sS)$   &$(RS)$       &$(RQ)$    &$(s\overline{Q}S)$ & $(sS)$   &$(RS)$    &$(RQ)$ &$(s\overline{Q}S)$&$(sS)$   &$(RS)$       &$(RQ)$ \\
\colrule
100   &0.93 & 1.19   & 1.66     & 2.40    &3.24 &2.60 &6.22 &11.65  &1.64 &1.29 &1.45 &1.34   &3.24 & 2.99    & 5.48  &11.21 \\
500   &0.70 & 0.87    & 1.21     & 1.49   &3.94 &2.77 &6.26 &11.74 &1.40 &1.06 &1.17 &1.15   &3.39 & 2.80  & 5.41  &11.14 \\
1000   &0.56 & 0.64    & 1.21     & 1.44  &3.92&2.70 &6.20 &11.59  &0.65 &0.75 &1.19 &1.35    &2.53 & 2.65  & 5.36  &10.97 \\
1500   &0.55 & 0.65    & 1.32     & 1.52  &4.09 &2.84 &6.24 &11.78  &0.52 &0.72 &1.31 &1.35   &2.35 & 2.63   & 5.37  &11.21 \\
\botrule
\end{tabular}}\label{table:stabilitytest}
\end{table}

The results of stability test show when the number of scenarios is 1000 or more, both mean STD and mean RMSE of genetic algorithm for the four controlling policies tend to be stable. Therefore, we select 1000 scenarios as the number of samples for genetic algorithm to solve our problem.

\subsection{Results and discussion}

In this section we present an extensive comparison of SDP, simulation-genetic algorithm and simulation-optimization heuristic pivoting on the test bed. Results are given by Table \ref{tab:pivot table1} and Table \ref{tab:pivot table2}.  As performance criterion, we adopt root mean square error (RMSE) and mean absolute percentage error (MAPE) between the optimal capital increment and the simulated capital increment under each given policy. Both RMSE and MAPE are employed because RMSE does not show as clearly as MAPE on how close simulated results are to optimal, while MAPE may result in some high percentage errors when optimal capital increments are near 0 even though optimal value and simulated values are close. 
The simulation confidence level is 95\%; details about the confidence intervals in the computational study for RMSE and MAPE are provided in Appendix 1.

\begin{table}[!ht]
\tbl{Pivot table for the computational study---RMSE.}
{\begin{tabular}[c]{{m{1.8cm}}*{5}{p{1.5cm}<{\centering}}{m{0.8cm}}} 
\toprule
\multirow{2}*{}&\multicolumn{5}{c}{Root Mean Square Error (RMSE)}&\multirow{3}*{Cases}\\
\cmidrule(lr){2-6}
&Sim-opt &GA-$s\overline{Q}S$&GA-$sS$ &GA-$RS$&GA-$RQ$\\ 
 \colrule
Ini capital &&&&&&\\
\specialrule{0em}{1pt}{1pt}
0 &14.30 &2.31&2.23 &6.64&16.20 &320\\
20 &12.80 &3.96&3.88 &5.89  &15.60&320\\
\specialrule{0em}{2pt}{2pt}
Interest rate &&&&&&\\
\specialrule{0em}{1pt}{1pt}
0.05 &13.60 &3.24 &3.22  &6.24  &15.40  &320\\ 
0.2  &13.61 &3.25 &3.11 &6.31  &16.40  &320\\
\specialrule{0em}{2pt}{2pt}
Selling price &&&&&&\\
\specialrule{0em}{1pt}{1pt}
5  &11.70 &2.36&2.22  &5.40 &14.30&320\\ 
10 &15.30 &3.94&3.89 &7.04 &17.40&320\\
\specialrule{0em}{2pt}{2pt}
Fixed order cost &&&&&&\\
\specialrule{0em}{1pt}{1pt}
10 &11.00 &2.87  &2.83 &6.26 &16.40&320\\ 
15 &15.80 &3.59 &3.47&6.29 &15.40&320\\
\specialrule{0em}{2pt}{2pt}
Vari order cost &&&&&&\\
\specialrule{0em}{1pt}{1pt}
1 &13.70 &3.31&3.29 &6.24 &15.10&320\\ 
2 &13.50 &3.18 &3.04 &6.32 &15.70&320\\
\specialrule{0em}{2pt}{2pt}
Penalty cost &&&&&&\\
\specialrule{0em}{1pt}{1pt}
2 &13.90 &3.79 &3.68 &5.46 &15.40&320\\ 
4 &13.30 &2.58 &2.56 &6.99 &16.40&320\\
\specialrule{0em}{2pt}{2pt}
Demand pattern &&&&&&\\
\specialrule{0em}{1pt}{1pt}
EMP1 &16.10 &3.01 &4.68 &5.13 &15.10&64\\ 
EMP2 &15.40 &3.19 &4.32 &5.63 &15.20&64\\
EMP3 &8.18  &2.83 &3.69 &5.43 &15.50&64\\ 
EMP4 &16.10 &3.33 &2.19 &7.28 &15.50&64\\
LCY1 &16.40 &1.81 &2.75 &8.64 &16.00&64\\ 
LCY2 &18.60 &1.74&3.26 &4.44 &15.20&64\\
RAND &11.40 &4.37 &1.64 &8.57 &15.50&64\\ 
SIN1 &10.60 &4.06 &1.82 &8.78 &15.50&64\\
SIN2 &12.20 &4.55&3.19 &7.38 &15.70&64\\ 
STA  &4.06 &2.04 &2.84 &7.11 &15.60&64\\
\specialrule{0em}{3pt}{3pt}
General &13.60&3.25&3.17&6.28&15.90&640\\
\specialrule{0em}{3pt}{3pt}
Aver. Time(s) &0.04&194.39&184.88&42.31&46.91&\\
 \botrule
\end{tabular}}\label{tab:pivot table1}
\end{table}

\begin{table}[!ht]
\tbl{Pivot table for the computational study---MAPE.}
{\begin{tabular}[c]{{m{1.8cm}}*{5}{p{1.5cm}<{\centering}}{m{0.8cm}}} 
\toprule
\multirow{2}*{}&\multicolumn{5}{c}{Mean Absolute Percentage Error (MAPE)}&\multirow{3}*{Cases}\\
\cmidrule(lr){2-6}
&Sim-opt &GA-$s\overline{Q}S$&GA-$sS$ &GA-$RS$&GA-$RQ$\\ 
 \colrule
Ini capital &&&&&&\\
\specialrule{0em}{1pt}{1pt}
0   &67.31\% &4.42\%  &4.78\%  &31.43\% &74.77\%&320\\
 20 &40.04\% &6.95\%  &6.39\%  &22.02\% &58.49\%&320\\
\specialrule{0em}{2pt}{2pt}
Interest rate &&&&&&\\
\specialrule{0em}{1pt}{1pt}
 0.05 &44.57\% &4.84\%  &5.02\% &22.42\% &50.05\% &320\\ 
0.2  &62.78\%  &6.53\%  &6.16\% &31.02\% &83.22\%&320\\
\specialrule{0em}{2pt}{2pt}
Selling price &&&&&&\\
\specialrule{0em}{1pt}{1pt}
5  &98.88\%&9.37\%   &9.16\%&49.39\%&123.30\%&320\\ 
10 &8.47\%&2.00\%   &2.02\% &4.06\% &9.97\%  &320\\
\specialrule{0em}{2pt}{2pt}
Fixed order cost &&&&&&\\
\specialrule{0em}{1pt}{1pt}
10 &33.64\%  &3.37\%  &3.93\% &15.51\%&42.06\%&320\\ 
15 &73.71\%  &8.00\% &7.25\%  &37.94\%&91.20\%&320\\
\specialrule{0em}{2pt}{2pt}
Vari order cost &&&&&&\\
\specialrule{0em}{1pt}{1pt}
1 &16.14\% &2.77\%  &2.8\%&7.77\%&19.06\%&320\\ 
2 &91.21\% &8.60\%  &8.37\%&45.67\%&114.20\% &320\\
\specialrule{0em}{2pt}{2pt}
Penalty cost &&&&&&\\
\specialrule{0em}{1pt}{1pt}
2 &34.43\%&5.00\%  &4.91\%&15.15\%&40.17\%&320\\ 
4 &72.92\% &6.37\%  &6.27\%&38.30\%&93.09\%&320\\
\specialrule{0em}{2pt}{2pt}
Demand pattern &&&&&&\\
\specialrule{0em}{1pt}{1pt}
EMP1 &2.52\%   &1.76\%  &1.68\%&6.06\%&19.88\%&64\\ 
EMP2 &104.25\% &10.29\%  &9.67\%&18.82\%&63.27\%&64\\
EMP3 &26.38\% &2.44\%  &2.25\%&11.18\%&31.43\%&64\\ 
EMP4 &6.82\%  &2.39\%  &2.20\%&8.19\%&21.95\%&64\\
LCY1 &92.92\% &5.15\%  &4.51\%&61.07\%&123.76\%&64\\ 
LCY2 &31.74\%&7.00\%  &5.00\%&89.91\%&193.40\%&64\\
RAND &29.04\%  &5.14\%  &6.01\%&5.48\%&23.29\%&64\\ 
SIN1 &39.55\%  &7.72\%  &8.08\%&10.54\%&35.18\%&64\\
SIN2 &23.44\% &8.02\%  &6.73\%&11.13\%&46.73\%&64\\ 
STA  &180.06\%&6.99\%  &9.56\%&46.15\%&107.43\%&64\\
\specialrule{0em}{3pt}{3pt}
General&53.67\%&5.68\%&5.59\%&26.72\%&66.63\%&640\\
\specialrule{0em}{3pt}{3pt}
Aver. Time(s) &0.04&194.39&184.88&42.31&46.91&\\
 \botrule
\end{tabular}}\label{tab:pivot table2}
\end{table}

Table \ref{tab:pivot table1} and Table \ref{tab:pivot table2} both show that policies $(s, S)$ and $(s, \overline{Q}, S)$ solved by genetic algorithm, in general perform better than other approaches (RMSE: 3.17 and 3.25, respectively;  MAPE: 5.68\% and 5.59\%, respectively), followed by policy $(R, S)$ (RMSE: 6.28, MAPE: 26.72\%), simulation-optimization heuristic (RMSE: 13.60, MAPE: 53.67\%) and policy $(R, Q)$ (RMSE: 15.90, MAPE: 66.63\%). Considering the confidence levels, performance of policy $(s, S)$ and policy $(s, \overline{Q}, S)$ are essentially identical. For the four controlling policies, it can be concluded that their performance is related with their flexibility. Since policy $(s, S)$ and policy $(s, \overline{Q}, S)$ are based on "dynamic uncertainty" strategy, which is most flexible, they perform best for the problem, while the least flexible policy $(R, Q)$ has worst performance. It is however surprising that enforcing a maximum order quantity $\overline{Q}$ does not seem to be beneficial, and that an $(s, S)$ policy with parameters carefully selected seems to provide competitive performances. Conversely, a simulation-optimization heuristic similar to that originally proposed by \cite{silver1973heuristic} --- which proved to be effective when the item is perishable \citep{gutierrez2017simple} --- performs poorly under capital flow.

The performance of different approaches does not seem to be affected by different parameter levels under the criterion RMSE; however, it is affected by the margin of product --- selling price and unit variable ordering cost under the MAPE criterion. Finally, the performance of the simulation-optimization heuristic varies substantially across different demand patterns.

In terms of computation times, the simulation-optimization heuristic runs faster than genetic algorithm, with average computation time less than one second (0.04s). Among the policies solved via genetic algorithm, policy $(R, S)$ runs fastest (42.31s), followed by policy $(R, Q)$ (46.91s), policy $(s, S)$ (184.88s), policy $(s, \overline{Q}, S)$ (194.39s). 








\section{Conclusion}

In this paper, we discussed a stochastic lot sizing model considering capital flow and business overdraft. In addition to obtaining optimal solutions via stochastic dynamic programming, we presented four controlling policies whose parameters have been computed via genetic algorithm, and a simulation-optimization heuristic. Our computational study demonstrate that policies $(s, S)$ and $(s, \overline{Q}, S)$ feature the smallest RMSE and MAPE with respect to the optimal solutions obtained by stochastic dynamic programming, followed by policy $(R, S)$, simulation-optimization heuristic and policy $(R, Q)$. Simulation-optimization heuristic is, however, the fastest approach. Numerical examples illustrate that capital availability, sales contribution margin and interest rates may substantially impact the optimal policy structure. These factors should be therefore taken into account by managers when making lot sizing decisions. 

Future research could proceed in several directions. When banks provide business overdraft, there is usually an upper limit for the borrowing capital in each period or floating interest rates for different amounts of borrowing capital; this is an extension that is worth investigating. Other extensions may consider different financing behaviors, such as trade credit, long term loans, and leasing. Finally, the optimal policy structure under capital flow constraints may also be explored.

\bibliographystyle{tPRS}
\bibliography{tPRSguide}

\begin{thebibliography}{33}
\newcommand{\enquote}[1]{``#1''}
\providecommand{\natexlab}[1]{#1}
\providecommand{\url}[1]{\normalfont{#1}}
\providecommand{\urlprefix}{ }
\expandafter\ifx\csname urlstyle\endcsname\relax
  \providecommand{\doi}[1]{doi:\discretionary{}{}{}#1}\else
  \providecommand{\doi}{doi:\discretionary{}{}{}\begingroup
  \urlstyle{rm}\Url}\fi

\bibitem[Askin(1981)]{askin1981}
Askin, Ronald~G. 1981.
\newblock ``A procedure for production lot sizing with probabilistic dynamic
  demand.'' \emph{Aiie Transactions} 13 (2): 132--137.

\bibitem[Axs{\"a}ter(2007)]{Axsater2007}
Axs{\"a}ter, Sven. 2007.
\newblock \emph{Inventory control}. Vol.~90. Springer Science \& Business
  Media.

\bibitem[Bollapragada and Morton(1999)]{bm1999}
Bollapragada, Srinivas, and Thomas~E Morton. 1999.
\newblock ``A simple heuristic for computing nonstationary {(s, S)} policies.''
  \emph{Operations Research} 47 (4): 576--584.

\bibitem[Bookbinder and Tan(1988)]{bookbinder1988strategies}
Bookbinder, James~H, and Jin-Yan Tan. 1988.
\newblock ``Strategies for the probabilistic lot-sizing problem with
  service-level constraints.'' \emph{Management Science} 34 (9): 1096--1108.

\bibitem[Bradley(2000)]{bradley2000lack}
Bradley, D. 2000.
\newblock ``Lack of financial and location planning causes small business
  bankruptcy.'' PhD diss., University of Central Arkansas, Conway, AR.

\bibitem[Buzacott and Zhang(2004)]{buzacott2004inventory}
Buzacott, John~A, and Rachel~Q Zhang. 2004.
\newblock ``Inventory management with asset-based financing.'' \emph{Management
  Science} 50 (9): 1274--1292.

\bibitem[Chao, Chen, and Wang(2008)]{chao2008dynamic}
Chao, Xiuli, Jia Chen, and Shouyang Wang. 2008.
\newblock ``Dynamic inventory management with cash flow constraints.''  .

\bibitem[Chen and Zhang(2015)]{Chen161}
Chen, Zhen, and Renqian Zhang. 2015.
\newblock ``A Single Item Lot Sizing Problem Considering Capital Flow and Trade
  Credit.'' \url{https://ssrn.com/abstract=2716525/}.

\bibitem[Coughtrie, Morley, and Ward(2009)]{coughtrie2009restructuring}
Coughtrie, D, J~Morley, and T~Ward. 2009.
\newblock ``Restructuring in bankruptcy: recent national case examples.''
  \url{https://www.eurofound.europa.eu/sites/default/files/ef_files/docs/erm/tn0908026s/tn0908026s.pdf/}.

\bibitem[Dorsey and Mayer(1995)]{dorsey1995genetic}
Dorsey, Robert~E, and Walter~J Mayer. 1995.
\newblock ``Genetic algorithms for estimation problems with multiple optima,
  nondifferentiability, and other irregular features.'' \emph{Journal of
  Business \& Economic Statistics} 13 (1): 53--66.

\bibitem[Elston and Audretsch(2011)]{elston2011financing22}
Elston, Julie~A, and David~B Audretsch. 2011.
\newblock ``Financing the entrepreneurial decision: an empirical approach using
  experimental data on risk attitudes.'' \emph{Small business economics} 36
  (2): 209--222.

\bibitem[Federgruen and Zipkin(1984)]{federgruenandzipkin1984}
Federgruen, Awi, and Paul Zipkin. 1984.
\newblock ``An efficient algorithm for computing optimal {(s, S)} policies.''
  \emph{Operations research} 32 (6): 1268--1285.

\bibitem[Gong, Chao, and Simchi-Levi(2014)]{gong2014dynamic}
Gong, Xiting, Xiuli Chao, and David Simchi-Levi. 2014.
\newblock ``Dynamic inventory control with limited capital and short-term
  financing.'' \emph{Naval Research Logistics (NRL)} 61 (3): 184--201.

\bibitem[Guan et~al.(2006)Guan, Ahmed, Nemhauser, and Miller]{guan2006branch}
Guan, Yongpei, Shabbir Ahmed, George~L Nemhauser, and Andrew~J Miller. 2006.
\newblock ``A branch-and-cut algorithm for the stochastic uncapacitated
  lot-sizing problem.'' \emph{Mathematical Programming} 105 (1): 55--84.

\bibitem[Gutierrez-Alcoba et~al.(2017)Gutierrez-Alcoba, Rossi, Martin-Barragan,
  and Hendrix]{gutierrez2017simple}
Gutierrez-Alcoba, Alejandro, Roberto Rossi, Belen Martin-Barragan, and
  Eligius~MT Hendrix. 2017.
\newblock ``A simple heuristic for perishable item inventory control under
  non-stationary stochastic demand.'' \emph{International Journal of Production
  Research} 55 (7): 1885--1897.

\bibitem[Ipsos(2017)]{IpsosMORI2017}
Ipsos, MORI. 2017.
\newblock ``2016 Business Finance Survey: SMEs.''
  \url{http://british-business-bank.co.uk/wp-content/uploads/2017/02/British-Business-Bank-Business-Finance-Survey-2016.pdf/}.

\bibitem[Kaut et~al.(2007)Kaut, Vladimirou, Wallace, and
  Zenios]{kaut2007stability}
Kaut, Michal, Hercules Vladimirou, Stein~W Wallace, and Stavros~A Zenios. 2007.
\newblock ``Stability analysis of portfolio management with conditional
  value-at-risk.'' \emph{Quantitative Finance} 7 (4): 397--409.

\bibitem[Koca, Yaman, and Akt{\"u}rk(2015)]{koca2015stochastic}
Koca, Esra, Hande Yaman, and M~Selim Akt{\"u}rk. 2015.
\newblock ``Stochastic lot sizing problem with controllable processing times.''
  \emph{Omega} 53: 1--10.

\bibitem[{\"O}zen, Do{\u{g}}ru, and Tarim(2012)]{ozen2012static}
{\"O}zen, Ula{\c{s}}, Mustafa~K Do{\u{g}}ru, and S~Armagan Tarim. 2012.
\newblock ``Static-dynamic uncertainty strategy for a single-item stochastic
  inventory control problem.'' \emph{Omega} 40 (3): 348--357.

\bibitem[Rossi, Kilic, and Tarim(2015)]{rossi2015piecewise}
Rossi, Roberto, Onur~A Kilic, and S~Armagan Tarim. 2015.
\newblock ``Piecewise linear approximations for the static--dynamic uncertainty
  strategy in stochastic lot-sizing.'' \emph{Omega} 50: 126--140.

\bibitem[Rossi et~al.(2014)Rossi, Tarim, Prestwich, and
  Hnich]{rossi2014piecewise}
Rossi, Roberto, S~Armagan Tarim, Steven Prestwich, and Brahim Hnich. 2014.
\newblock ``Piecewise linear lower and upper bounds for the standard normal
  first order loss function.'' \emph{Applied Mathematics and Computation} 231:
  489--502.

\bibitem[Scarf(1959)]{scarf1959}
Scarf, Herbert. 1959.
\newblock ``The optimality of {(s, S)} policies in the dynamic inventory
  problem.''  .

\bibitem[Sexton, Dorsey, and Johnson(1998)]{sexton1998toward}
Sexton, Randall~S, Robert~E Dorsey, and John~D Johnson. 1998.
\newblock ``Toward global optimization of neural networks: a comparison of the
  genetic algorithm and backpropagation.'' \emph{Decision Support Systems} 22
  (2): 171--185.

\bibitem[Silver(1978)]{silver1978inventory}
Silver, Edward. 1978.
\newblock ``Inventory control under a probabilistic time-varying, demand
  pattern.'' \emph{Aiie Transactions} 10 (4): 371--379.

\bibitem[Silver and Meal(1973)]{silver1973heuristic}
Silver, Edward~A, and Harlan~C Meal. 1973.
\newblock ``A heuristic for selecting lot size quantities for the case of a
  deterministic time-varying demand rate and discrete opportunities for
  replenishment.'' \emph{Production and inventory management} 14 (2): 64--74.

\bibitem[Sophie~Doove(2014)]{Survey2014}
Sophie~Doove, Ton Kwaak Lia Smit Tommy~Span, Petra~Gibcus. 2014.
\newblock ``Survey on the access to finance of enterprises.''
  \url{http://wwwe.ansa.it/documents/1415814222451_Rapporto.pdf/}.

\bibitem[Tarim and Kingsman(2004)]{tarim2004stochastic}
Tarim, S~Armagan, and Brian~G Kingsman. 2004.
\newblock ``The stochastic dynamic production/inventory lot-sizing problem with
  service-level constraints.'' \emph{International Journal of Production
  Economics} 88 (1): 105--119.

\bibitem[Tarim and Kingsman(2006)]{tarim2006modelling}
Tarim, S~Armagan, and Brian~G Kingsman. 2006.
\newblock ``Modelling and computing (R n, S n) policies for inventory systems
  with non-stationary stochastic demand.'' \emph{European Journal of
  Operational Research} 174 (1): 581--599.

\bibitem[Tempelmeier(2007)]{tempelmeier2007stochastic}
Tempelmeier, Horst. 2007.
\newblock ``On the stochastic uncapacitated dynamic single-item lotsizing
  problem with service level constraints.'' \emph{European Journal of
  Operational Research} 181 (1): 184--194.

\bibitem[Tempelmeier and Herpers(2011)]{tempelmeier2011dynamic}
Tempelmeier, Horst, and Sascha Herpers. 2011.
\newblock ``Dynamic uncapacitated lot sizing with random demand under a
  fillrate constraint.'' \emph{European Journal of Operational Research} 212
  (3): 497--507.

\bibitem[Wagner and Whitin(1958)]{wagner1958dynamic}
Wagner, Harvey~M, and Thomson~M Whitin. 1958.
\newblock ``Dynamic version of the economic lot size model.'' \emph{Management
  science} 5 (1): 89--96.

\bibitem[Wuttke et~al.(2016)Wuttke, Blome, Heese, and
  Protopappa-Sieke]{wuttke2016supply}
Wuttke, David~A, Constantin Blome, H~Sebastian Heese, and Margarita
  Protopappa-Sieke. 2016.
\newblock ``Supply chain finance: Optimal introduction and adoption
  decisions.'' \emph{International Journal of Production Economics} 178:
  72--81.

\bibitem[Zeballos, Seifert, and Protopappa-Sieke(2013)]{zeballos2013single}
Zeballos, Ariel~C, Ralf~W Seifert, and Margarita Protopappa-Sieke. 2013.
\newblock ``Single product, finite horizon, periodic review inventory model
  with working capital requirements and short-term debt.'' \emph{Computers \&
  Operations Research} 40 (12): 2940--2949.

\end{thebibliography}

\newpage
\appendices
\section{Confidence levels for the computational study.}
Table \ref{tab:confidence levels1} provides details of confidence levels in our computational study under criterion RMSE and Table \ref{tab:confidence levels2} gives the confidence intervals under criterion MAPE.

\begin{table}[!ht]
\tbl{Details of confidence levels for the computational study---RMSE.}
{\begin{tabular}[c]{{p{1.8cm}}*{5}{p{1.5cm}<{\centering}}{p{0.8cm}}}
\toprule
\multirow{2}*{}&\multicolumn{5}{c}{0.95 Confidence intervals---RMSE}&\multirow{2}*{Cases}\\
\cmidrule(lr){2-6}
&Sim-opt &GA-sQS&GA-sS &GA-RS&GA-RQ \\
 \colrule
Ini capital &&&&&&\\
\specialrule{0em}{1pt}{1pt}
0  &[13.30,15.60] &[2.15,2.50] &[2.07,2.42] &[6.16,7.20] &[15.08,17.60] &320\\
20  &[11.89,13.90]&[3.68,4.29] &[3.60,4.21] &[5.47,6.38] &[15.46,16.90]&320\\
\specialrule{0em}{2pt}{2pt}
Interest rate &&&&&&\\
\specialrule{0em}{1pt}{1pt}
0.05  &[12.61,14.70]&[3.01,3.51]  &[2.99,3.49] &[5.79,6.76] &[14.34,16.70] &320\\ 
0.2  &[12.63,14.80]&[3.02,3.52] &[2.89,3.37] &[5.86,6.84] &[15.19,17.70] &320\\
\specialrule{0em}{2pt}{2pt}
Selling price &&&&&&\\
\specialrule{0em}{1pt}{1pt}
5  &[10.85,12.70]&[2.19,2.56] &[2.06,2.41] &[5.01,5.85]&[13.29,15.50]&320\\ 
10 &[14.17,16.60]&[3.65,4.27] &[3.61,4.21]&[6.53,7.63]&[16.11,18.80]&320\\
\specialrule{0em}{2pt}{2pt}
Fixed order cost &&&&&&\\
\specialrule{0em}{1pt}{1pt}
10 &[10.23,11.90]&[2.66,3.11] &[2.63,3.07]&[5.81,6.78]&[15.20,17.70]&320\\ 
15 &[14.63,17.10]&[3.33,3.89] &[3.22,3.76]&[5.84,6.82]&[14.33,16.70]&320\\
\specialrule{0em}{2pt}{2pt}
Vari order cost &&&&&&\\
\specialrule{0em}{1pt}{1pt}
1 &[12.68,14.80]&[3.08,3.59] &[3.05,3.57]&[5.79,6.76]&[14.04,16.40]&320\\ 
2 &[12.65,14.70]&[2.95,3.44] &[2.82,3.29]&[5.86,6.85]&[16.46,18.10]&320\\
\specialrule{0em}{2pt}{2pt}
Penalty cost &&&&&&\\
\specialrule{0em}{1pt}{1pt}
2 &[12.86,15.00]&[3.52,4.11] &[3.41,3.98]&[5.07,5.92]&[14.34,16.70]&320\\ 
4 &[12.38,14.50]&[2.40,2.80] &[2.37,2.77]&[6.49,7.58]&[15.19,17.70]&320\\
\specialrule{0em}{2pt}{2pt}
Demand pattern &&&&&&\\
\specialrule{0em}{1pt}{1pt}
EMP1 &[13.76,19.50]&[2.57,3.64] &[3.99,5.66]&[4.37,6.20]&[12.90,18.30]&64\\ 
EMP2 &[13.12,18.60]&[2.73,3.86] &[3.68,5.22] &[4.80,6.81]&[12.92,18.30]&64\\
EMP3 &[6.98,9.90]&[2.41,3.42] &[3.15,4.46] &[4.63,6.57]&[13.22,18.70]&64\\ 
EMP4 &[13.75,19.50]&[2.84,4.03] &[1.87,2.65]&[6.21,8.80]&[12.95,18.40]&64\\
LCY1 &[14.01,19.90]&[1.54,2.18] &[2.35,3.33]&[7.37,10.40]&[13.67,19.40]&64\\ 
LCY2 &[15.88,22.50]&[1.49,2.11] &[2.78,3.94]&[3.78,5.37]&[12.94,18.30]&64\\
RAND &[9.73,13.80]&[3.73,5.29] &[1.39,1.98] &[7.31,10.40]&[13.25,18.80]&64\\ 
SIN1 &[9.05,12.80]&[3.46,4.91] &[1.55,2.20]&[7.49,10.60]&[13.20,18.87]&64\\
SIN2 &[10.42,14.80]&[3.88,5.50] &[2.72,3.86]&[6.29,8.92]&[13.42,19.00]&64\\ 
STA  &[3.47,4.91]&[1.74,2.47] &[2.42,3.43]&[6.06,8.60]&[13.58,19.30]&64\\
\specialrule{0em}{3pt}{3pt}
General &[12.89,14.40]&[3.07,3.44] &[3.00,3.35]&[5.95,6.64]&[15.09,16.80]&640\\
 \botrule
\end{tabular}}\label{tab:confidence levels1}
\end{table}

\newpage
\begin{table}[!ht]
\tbl{Details of confidence levels for the computational study---MAPE.}
{\begin{tabular}[c]{{p{1.8cm}}*{5}{p{1.5cm}<{\centering}}{p{0.8cm}}}
\toprule
\multirow{2}*{}&\multicolumn{5}{c}{0.95 Confidence intervals---MAPE}&\multirow{2}*{Cases}\\
\cmidrule(lr){2-6}
&Sim-opt &GA-sQS&GA-sS &GA-RS&GA-RQ \\
 \colrule
Ini capital &&&&&&\\
\specialrule{0em}{1pt}{1pt}
0 &$\pm$36.63\% &$\pm$1.26\% &$\pm$2.00\%  &$\pm$15.36\% &$\pm$32.11\%&320\\
20 &$\pm$14.58\% &$\pm$1.87\% &$\pm$1.39\% &$\pm$12.49\% &$\pm$32.08\%&320\\
\specialrule{0em}{2pt}{2pt}
Interest rate &&&&&&\\
\specialrule{0em}{1pt}{1pt}
0.05  &$\pm$16.44\% &$\pm$1.52\% &$\pm$1.13\%&$\pm$10.62\%&$\pm$17.70\%&320\\ 
0.2   &$\pm$35.86\% &$\pm$1.94\% &$\pm$2.17\%&$\pm$16.69\%&$\pm$41.74\%&320\\
\specialrule{0em}{2pt}{2pt}
Selling price &&&&&&\\
\specialrule{0em}{1pt}{1pt}
5  &$\pm$38.83\%&$\pm$2.19\% &$\pm$2.38\%&$\pm$19.49\% &$\pm$44.55\%&320\\ 
10 &$\pm$0.49\% &$\pm$0.11\% &$\pm$0.10\%&$\pm$0.18\%  &$\pm$0.33\% &320\\
\specialrule{0em}{2pt}{2pt}
Vari order cost &&&&&&\\
\specialrule{0em}{1pt}{1pt}
1 &$\pm$2.25\%  &$\pm$0.30\%  &$\pm$0.27\%&$\pm$0.94\%  &$\pm$1.65\%&320\\ 
2 &$\pm$38.98\% &$\pm$2.20\%  &$\pm$2.39\%&$\pm$19.57\% &$\pm$44.77\%&320\\
\specialrule{0em}{2pt}{2pt}
Penalty cost &&&&&&\\
\specialrule{0em}{1pt}{1pt}
2 &$\pm$15.23\% &$\pm$1.05\% &$\pm$0.94\%&$\pm$8.92\%&$\pm$19.49\%&320\\ 
4 &$\pm$36.30\% &$\pm$2.01\% &$\pm$2.26\%&$\pm$17.59\%&$\pm$40.81\%&320\\
\specialrule{0em}{2pt}{2pt}
Demand pattern &&&&&&\\
\specialrule{0em}{1pt}{1pt}
EMP1 &$\pm$0.58\%  &$\pm$0.13\% &$\pm$0.15\% &$\pm$1.86\% &$\pm$4.23\%&64\\ 
EMP2 &$\pm$63.98\% &$\pm$5.53\% &$\pm$5.31\% &$\pm$10.09\%&$\pm$33.15\%&64\\
EMP3 &$\pm$8.11\% &$\pm$0.30\% &$\pm$0.21\%&$\pm$3.33\%&$\pm$8.90\%&64\\ 
EMP4 &$\pm$0.97\%  &$\pm$0.37\% &$\pm$0.34\%&$\pm$3.52\%&$\pm$5.96\%&64\\
LCY1 &$\pm$72.12\% &$\pm$3.04\% &$\pm$1.94\%&$\pm$48.27\%&$\pm$100.05\%&64\\ 
LCY2 &$\pm$14.39\% &$\pm$3.26\% &$\pm$2.00\%&$\pm$75.57\%&$\pm$170.44\%&64\\
RAND &$\pm$10.12\%  &$\pm$1.78\% &$\pm$2.32\%&$\pm$1.31\%&$\pm$8.03\%&64\\ 
SIN1 &$\pm$17.08\% &$\pm$5.98\% &$\pm$3.89\%&$\pm$4.35\%&$\pm$13.87\%&64\\
SIN2 &$\pm$9.49\%&$\pm$4.44\% &$\pm$3.16\%&$\pm$3.79\%&$\pm$17.49\%&64\\ 
STA  &$\pm$169.98\% &$\pm$4.38\% &$\pm$9.23\%&$\pm$40.71\%&$\pm$105.10\%&64\\
\specialrule{0em}{3pt}{3pt}
General&$\pm$19.69\%&$\pm$1.13\%&$\pm$1.22\%&$\pm$9.88\%&$\pm$22.64\%&640\\
 \botrule
\end{tabular}}\label{tab:confidence levels2}
\end{table}

\end{document}